\documentclass[preprint2]{aastex}

\newcommand{\simgt}{\,\rlap{\lower 3.5 pt \hbox{$\mathchar \sim$}} \raise
1pt \hbox {$>$}\,}
\newcommand{\simlt}{\,\rlap{\lower 3.5 pt \hbox{$\mathchar \sim$}} \raise
1pt \hbox {$<$}\,}

\shorttitle{Dark matter velocity anisotropy}
\shortauthors{Steen H. Hansen}

\begin{document}


\title{Might we eventually 
understand the origin of the dark matter velocity anisotropy?}


\author{Steen H. Hansen}
\affil{Dark Cosmology Centre, Niels Bohr Institute, University of Copenhagen,\\
Juliane Maries Vej 30, 2100 Copenhagen, Denmark}
\email{hansen@dark-cosmology.dk}



\begin{abstract}
The density profile of simulated dark matter structures is fairly
well-established, and several explanations for its characteristics
have been put forward. In contrast, the radial variation of the
velocity anisotropy has still not been explained. We suggest a very
simple origin, based on the shapes of the velocity distributions
functions, which are shown to differ between the radial and tangential
directions.  This allows us to derive 
a radial variation of the anisotropy profile which is
in good agreement with both simulations and observations.  One of the
consequences of this suggestion is that the velocity anisotropy is
entirely determined once the density profile is known. We demonstrate
how this explains the origin of the $\gamma$--$\beta$ relation, which
is the connection between the slope of the density profile and the
velocity anisotropy.  These findings provide us with a powerful tool,
which allows us to close the Jeans equations.
\end{abstract}




\section{Introduction}

The natural outcome of cosmological structure formation theory is
equilibrated dark matter (DM) structures. According to numerical
simulations, the mass density profile, $\rho(r)$, of these structures
changes from something with a fairly shallow profile in the central
region, $\gamma \equiv d{\rm ln}\rho/d{\rm ln}r \sim -1$ (or maybe
zero), to something steeper in the outer region, $\gamma \sim -3$ (or
maybe steeper) \citep{nfw,moore,diemand} (see also
\cite{reed,stoehr,navarro2004,alister,merritt,ascasibar,stadel08,springel08}).  For the
largest structures, like galaxy clusters, there appears to be fair
agreement between the numerical predictions and observations
concerning the central steepness
\citep{pointe,sand,buote,broadhurst,vikhlinin}, however, for smaller
structures, like galaxies or dwarf galaxies, observations tend to
indicate central cores~\citep{salucci,gilmore,wilkinson} (see
also~\cite{rubin85,courteau97,palunas00,blok01,blok02,salucci01,swaters02,corb03,salucci2}).
The various theoretical approaches still make different predictions
\citep{taylornavarro,hansenjeans,austin,dehnenmclaughlin,gsmh,henriksen,henriksen2},
varying from central cores to cusps.

The second natural quantity to consider (after the density
profile) is the velocity anisotropy, which is defined through
\begin{equation}
\beta \equiv 1 - \frac{\sigma^2_t}{\sigma^2_r} ~,
\end{equation}
where $\sigma^2_t$ and $\sigma^2_r$ are the 1-dimensional tangential
and radial velocity dispersions~\citep{binneytremaine}. If most dark
matter particles in an equilibrated structure were purely on radial
orbits, then $\beta$ could be as large as 1, and for mainly
tangential orbits $\beta$ could be arbitrarily large and
negative. Since dark matter is collision-less $\beta$ does not have to
be zero, and it could in principle even vary as a function of radius.

Numerical N-body simulations of collision-less dark matter particles
show that the dark matter velocity aniso\-tro\-py is indeed radially
varying, and that $\beta$ goes from roughly zero in the central region, to 0.5
towards the outer region~\citep{colelacey,carlberg}.  Only very
recently has this velocity anisotropy been measured to be non-zero in
galaxy clusters \citep{hansenpiff}, and it has even been observed to
be increasing as a function of radius \citep{host2008}, in excellent
agreement with the numerical predictions. For smaller structures, like
our own galaxy, this has not been observed yet. In principle $\beta$
of our Galaxy can be measured in an underground directional sensitive
detector, however, it will require a large de\-dicated experimental
programme~\citep{hosthansen}.  Very little theoretical understanding
of the origin of this velocity anisotropy exists, and to my knowledge
no successful derivation of it has been published (see,
however, \cite{hansenmoore,smgh07,wojtak2008}).  We will in this paper
present an attempt towards deriving $\beta$.

\section{Decomposition}

When analyzing the outcome of a numerically simulated dark matter
structure one traditionally divides the equilibrated structure in bins
(shells) in radius, or in potential energy. For spherical structures
there is naturally no difference. We can now consider all the
particles in a given radial bin, and calculate properties like average
density, angular momentum, velocity anisotropy etc. In order to do
this, we must decompose the velocity of each particle into the radial
component, and the two tangential components.  
The two tangential components can for instance
be separated according to the total angular momentum of all the
particles in the bin.

By summing over all the particles in the radial (or potential) bin, we
thus get the velocity distribution function (VDF), which for a gas
would have been a Gaussian represented by the local gas temperature,
$f(v) \sim {\rm exp} (-E/T)$.  We are here discussing the
1-dimensional VDF (i.e. the one where the two other velocities are
integrated over), and we are not assuming that the radial and
tangential VDF's are independent.  Naturally, since dark matter
particles are not collisional, the concept of temperature is not well
defined for them. In numerically simulated structures one observes
that the radial VDF is symmetric (with respect to particles moving in
or out of the structure), and also the non-rotational part of the
tangential VDF is symmetric. The asymmetry of the rotational part of
the tangential VDF was discussed in \cite{schmidt} for the DM
particles.  For the structures with very little rotation, the two
tangential VDF's are virtually identical.  To avoid any complications
from the total angular momentum we will hereafter only discuss the two
symmetric 1-dimensional VDF's.
For any given radial bin in a given DM structure, the shape of the VDF
only depends on the momentaneous distribution of particles (which should
be virtually time-independent for equilibrated structures), and is 
independent of the method by which the structure is selected.

\begin{figure}[thb]
	\centering
	\includegraphics[angle=0,width=0.49\textwidth]{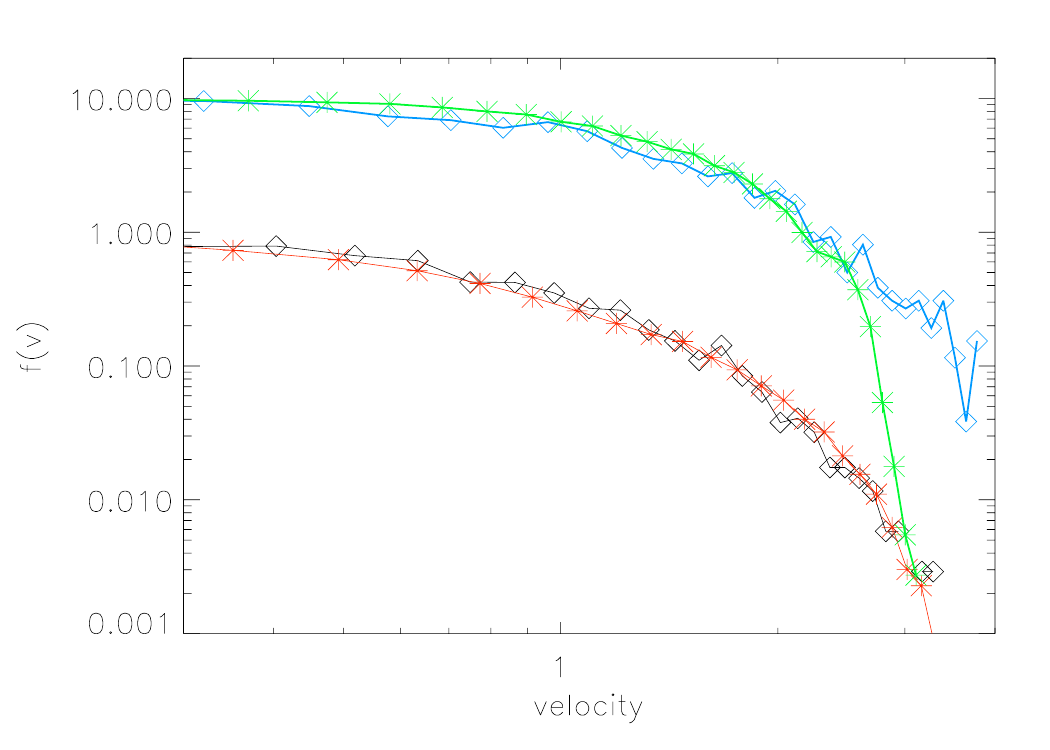}
	\caption{The velocity distribution function for 2 different radial bins 
	 from a simulated cosmological DM structure \citep{moore2001}.
The upper (green and blue) curves are the radial VDF, and the lower
(black and red) are the tangential VDF. The open diamonds are from the
inner bin, and stars are from the outer bin. It is clear that the
tangential VDF's are very similar to each other, whereas the radial
VDF's differ in shape both at small and large velocities.  All
figures have velocity normalized to the dispersion, and random y-axis
normalization to enhance visibility.}
\label{fig:vdf}
\end{figure}

When analyzing dark matter structures resulting from cosmological
simulations, we find that the shape of the {\em radial} VDF changes
as a function of radius~\citep{wojtak2005,hansenzemp,faltenb,fairbairn}. 
In particular, the bins in the inner region
tend to have long tails (more particles at high velocity compared to a
Gaussian), whereas bins at larger radii tend to have stronger
reduction in high velocity particles.  This is exemplified in figure
\ref{fig:vdf}, where the upper curves (blue and green) show the radial
VDF.  The open diamonds (blue) come from a radial bin in the inner
region, whereas the stars (green) are from a bin further out. 
The VDF's are normalized such that a comparison is possible, and 
velocity is normalized to the dispersion.
This
simulated cosmological data is from the Local Group
simulation of \cite{moore2001}. The lower curves (red and black) are
the tangential VDF from the same two bins, however, for the {\em
tangential} VDF there is a striking resemblance, infact to a first
approximation these two tangential VDF's from different radial bins
look identical.


\begin{figure}[thb]
	\centering
	\includegraphics[angle=0,width=0.49\textwidth]{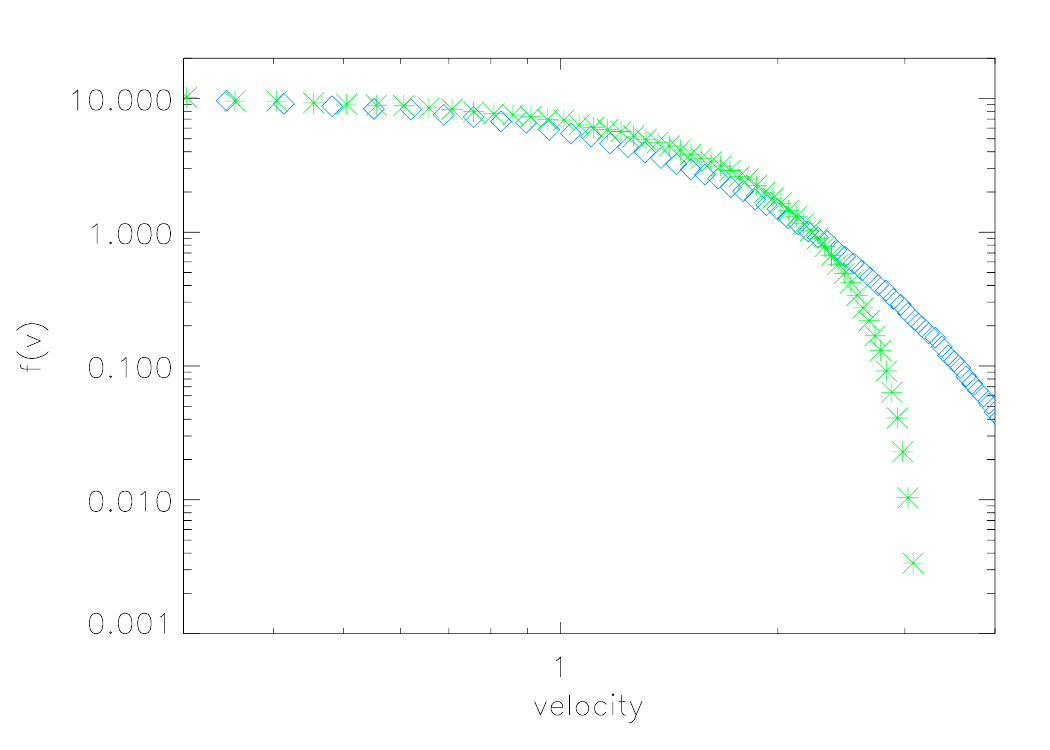}
	\caption{The velocity distribution function for 2 different radial bins
	 from the Eddington formula for an NFW density profile.  The
open diamonds are from an inner bin ($r=0.1$), and stars are from an
outer bin ($r=10$).  There is a striking resemblance with the radial
VDF's from the cosmological si\-mulation in the upper curves in figure \ref{fig:vdf}.
Same normalization as figure  \ref{fig:vdf}.}
\label{fig:edd}
\end{figure}


The most frequently used approach to discuss DM structures is through
the first Jeans equation, which relates the velocity dispersions to
the density profiles.  If we were to have some knowledge about some of the
quantities entering the Jeans equation, then we can solve for the
others. One example hereof was presented in~\cite{dehnenmclaughlin},
who demonstrated how to derive a generalized NFW density profile, by
both assuming that the pseudo phase-space is a power-law in radius
\citep{taylornavarro}, and that there is a linear relation between the
velocity anisotropy and the density slope \citep{hansenmoore}. A
somewhat generalized approach was presented in~\cite{zait}, where
the authors demonstrated how to derive the velocity anisotropy by
assuming simple forms for both the density profile and the pseudo
phase-space density. The fundamental problem with this kind of
approaches is, that any departure from truth in the assumptions will
lead to departure from correctness in the results. \cite{zait}
demonstrated this in a very convincing way, by deriving significantly
different velocity anisotropy profiles by just changing between NFW or
Sersic density profiles as input. Another related problem with these
approaches is, that the assumption that pseudo phase-space is a simple
power-law, was recently demonstrated to be oversimplified.  The
unknown question was which component (radial, tangential, or something
else) of the velocity dispersion in the pseudo phase-space gives the
best approximation to a power-law in radius
\citep{hansenzemp,knollmann}.  \cite{schmidt09} demonstrated that 
different numerically simulated structures are best fitted by
different forms of the pseudo phase-space, and hence that there is no
universal simple behavior of the pseudo phase-space.

\subsection{The shape of the radial VDF}

For ergodic structures, that is structures where the orbits depend
only on energy (hence with $\beta=0$) we can use the Eddington
formula to get the VDF at any radius~\citep{eddington}, (see
also~\cite{binney82,cuddeford,evansan}).  
The Eddington formula only
depends on the radial dependence of the density profile of the
structure, and by assumption the VDF is the same for the radial and
tangential directions, $f(v,r) = $ function$(\rho(r))$.
It is natural to interpret this  in
the following way.  The structure is in equilibrium, so there is detailed balance
for each phase-space element. The velocity of each particle is
decomposed into the radial and tangential components, and for any
infinitesimal time step, the radial component of any individual particle can tell
that it is moving in a changing density and changing potential (the radial
component of any individual particle is either moving directly inwards or
outwards).
It is
therefore natural that the radial VDF is imprinted by the radial
variation in the density and potential.  For a truncated NFW density
profile we can e.g. get the VDF at radius 0.1 and 10, in units of the
characteristic radius, see figure \ref{fig:edd}. By comparison of
figures \ref{fig:vdf} and \ref{fig:edd} it is clear, that the radial
VDF of the cosmological simulation indeed looks very similar to the
VDF from the Eddington formula.  It is therefore tempting to suggest
that the radial VDF to a first approximation is identical to 
the one which results when applying the
Eddington formula to the given density profile.

Now, the actual radial VDF (for a given cosmologically simulated
structure) will differ slightly from the VDF resulting from the
Eddington formula, since the latter was based on the assumption that
$\beta=0$. Specifically, the VDF from the Eddington formula gives also
$\sigma_r$, and to ensure consistency with the Jeans equation, one
must have $\beta=0$.  However, we will here use this VDF {\em as a
first approximation} to the radial VDF, and we will present a
quantitative comparison in a future paper.
Recently, \cite{vanhese} showed that for a large class of theoretical
model, this is an excellent approximation (see their figure 5).

\subsection{The shape of the tangential VDF}

It is somewhat less trivial to 
argue (or claim) the shape of the tangential VDF.
For an infinitesimal time step, the tangential component of any individual
particle's velocity is moving in constant density and constant potential (the tangential velocity
component of any individual particle is moving, well, tangentially, and we assume
spherical symmetry).  We still assume that the structure is in equilibrium
with no time variation. This
means, that {\em as a first approximation} the tangential VDF can be
thought of as the one resulting from an infinite and homogeneous
medium, where both the density and potential is constant
everywhere. This argument is similar to the Jeans swindle, where the
homogeneous medium implies constant potential.
Naturally, such an infinite structure is not
gravitationally stable against perturbations, but we can instead
approximate it in the following way.

Let us consider a density profile, which is a power-law in radius over
many orders of magnitude, and is then truncated.  One example hereof
is an NFW profile, where the central density slope is -1, and the
corresponding truncation is then happening after the scale radius. We
are thus considering the VDF in a bin at a radius which is many orders
of magnitude deeper towards the center than the scale radius. 
We can now consider a generalized double power-law profile, where the central
slope can be more shallow than -1, and we can
use the Eddington formula to extract the VDF for any central slope. 
By lowering the central slope towards zero, we  get
a structure which in principle is stable towards perturbations, but at
the same time is approaching constant density and constant
potential in the central region. The resulting VDF (extrapolated to zero slope) has been
discussed by \cite{zurichstudents} and has the shape
\begin{equation}
f(v) = n(\rho) \, \left( 1 -  \frac{1-q}{3-q} \, 
\left( \frac{v}{\sigma}  \right) ^2 \right) ^ \frac{q}{1-q}  \, ,
\label{eq:tsallis}
\end{equation}
with $q=5/3$, and $n(\rho)$ shows that the normalization only depends
on the local density. This form is known as a q-generalized exponential~\citep{tsallis}.
For comparison one should note that a simple comparison with 
polytropes (where 
$f(E) \sim E^{(n-3/2)}$) breaks down, since the normal connection 
between density and potential, $\rho \sim \Psi ^n$, is not
valid for such shallow slopes~\citep{binneytremaine}.


\begin{figure}[thb]
	\centering
	\includegraphics[angle=0,width=0.49\textwidth]{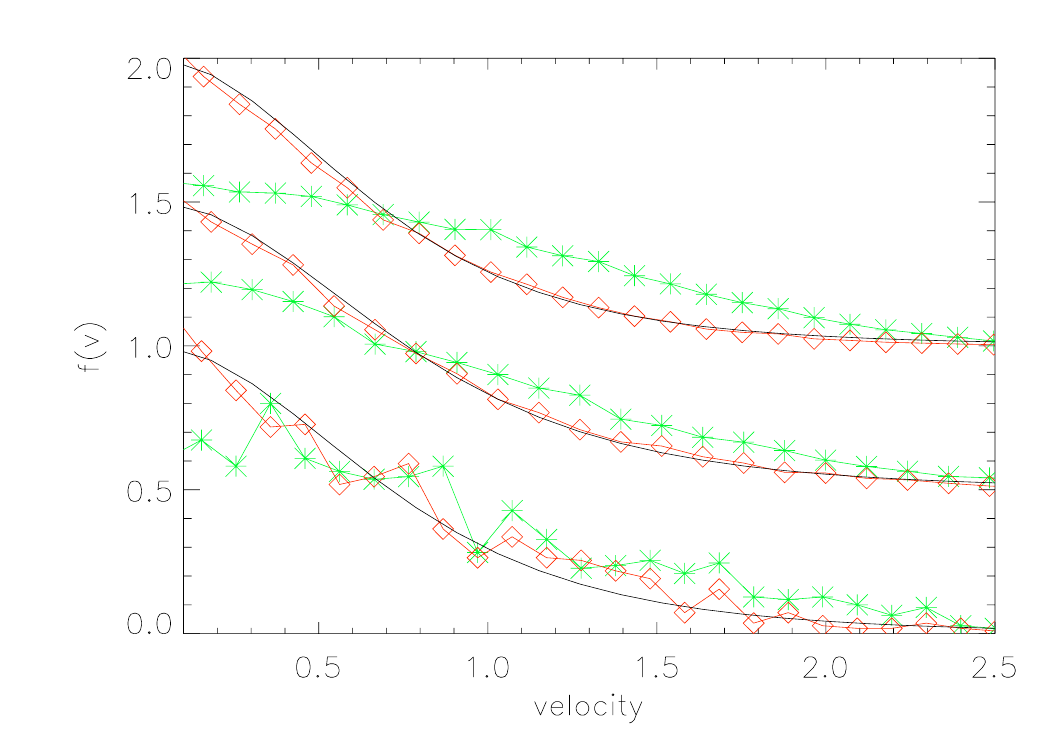}
	\caption{The VDF as function of velocity at 3 different radii,
upper curves correspond to outer radial bin, and lower curves
correspond to an inner radial bin. The green stars are the radial
VDF's, whereas the red diamonds are the tangential VDF's. The
simulation is the very non-cosmological ``tangential orbit
instability'' from~\cite{hansenzemp}.  The black lines are of the form
in eq.~(\ref{eq:tsallis}). The velocity is normalized to the dispersion, 
and the y-axis has been shifted vertically for two of the bins
to enhance visibility. One clearly sees that the tangential VDF's are virtually
identical, whereas the radial VDF's vary as function of radius.}
\label{fig:tangent}
\end{figure}

\begin{figure}[thb]
	\centering
	\includegraphics[angle=0,width=0.49\textwidth]{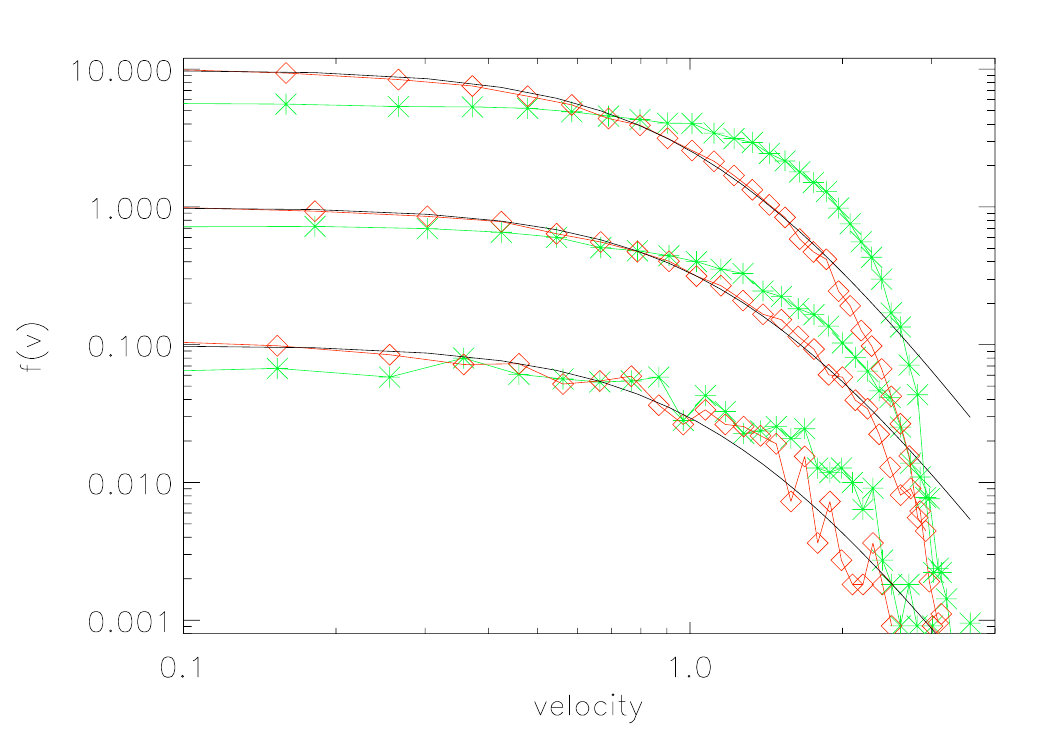}
	\caption{Same as fig.~\ref{fig:tangent} but with log-scales
to make the suppression at high-energy more visible. The normalization of
the y-axis is random, to enhance visibility.}
\label{fig:tangent.log}
\end{figure}


When considering the shape of the tangential VDF from simulations in figs.
\ref{fig:tangent} and \ref{fig:tangent.log} we see that this form
indeed provides a very good fit for all radii, at least for $v$
smaller than roughly $2\sigma$. The structure in
figs. \ref{fig:tangent} and \ref{fig:tangent.log} is from a very
non-cosmological simulation (the ``tangential orbit instability'' of
\cite{hansenzemp}).  The same form fits the tangential VDF's from a
cosmological simulation (lower lines in fig.~\ref{fig:vdf}) equally
well.

Clearly, this form in eq.~(\ref{eq:tsallis}) has an extended tail of
high energy particles, which would not be bound by the equilibrated
structure. The suppression of high energy particles due to the finite
radial extend of the structure is naturally included through the
Eddington formula for the radial component, and we therefore make
the suggestion that the {\em tangential} VDF must have a high-energy
tail which follows the {\em radial} VDF. Effectively, this means that
for large velocities the tangential
component of the velocity might as well be moving in the radial
direction. This corresponds to the fact that
the tangential velocity component of any individual particle actually 
is moving somewhat radially after a finite time-interval.

When looking at fig. \ref{fig:tangent.log} we see that the actual
suppression is even slightly larger for these high-energy particles,
however, the difference between the suggested and the actual
suppression at high energy is very small. When looking at the number
of particles (the integral under the curve in
fig. \ref{fig:tangent}) we find, that the difference is virtually
zero.

In conclusion, the tangential VDF is surprisingly well fit by the
phenomenologically predicted shape in eq.~(\ref{eq:tsallis}), and with
a high-energy tail suppression corresponding to the tail of the radial
VDF.

\begin{figure}[thb]
	\centering
	\includegraphics[angle=0,width=0.49\textwidth]{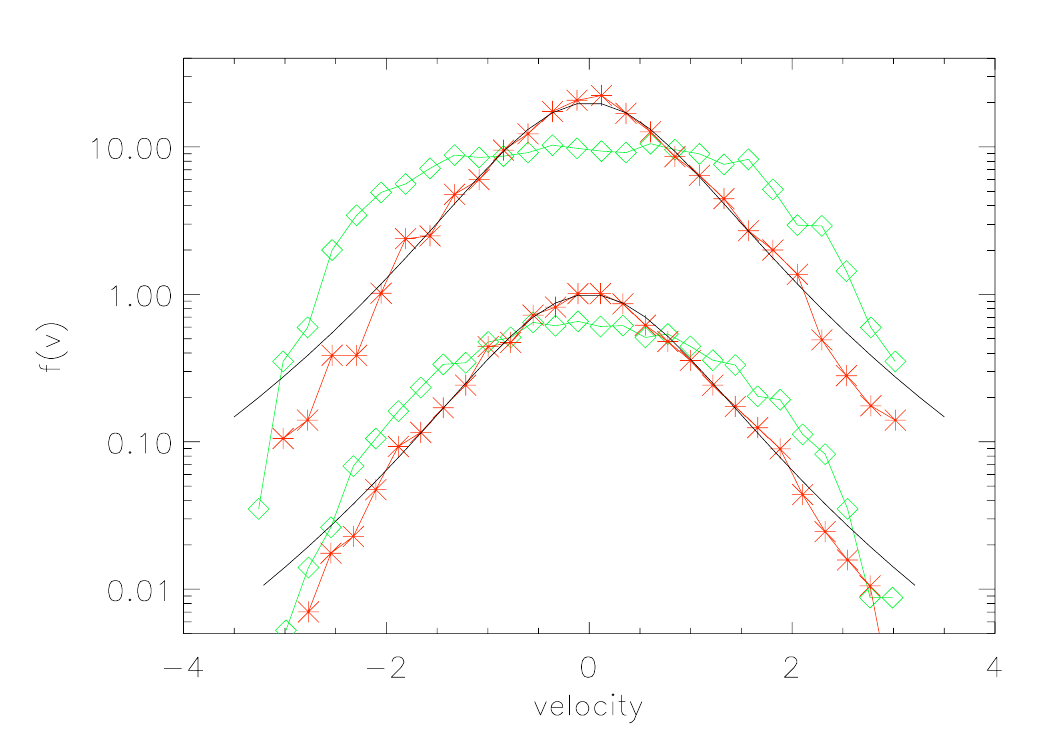}
	\caption{The velocity distribution function as function of velocity
for a galaxy from a cosmological simulation including both gas and stars
 from \cite{sommerlarsen2,sommerlarsen}. Green diamonds are radial VDF's,
red stars are tangential VDF's, and the lines are of the form in 
eq.~(\ref{eq:tsallis}).}
\label{fig:k15}
\end{figure}

To emphasize the general nature of the shape of the tangential VDF, we
also present the radial and tangential VDF's of a cosmological
si\-mulation of a galaxy, including both cooling gas, star-formation and
stars, as well as
supernova feedback
from \cite{sommerlarsen2,sommerlarsen}. 
In fig.~\ref{fig:k15} we see that the dark matter VDF
also in this case has the suggested shape.


\section{The velocity anisotropy}
Now, after having established the shape of both the radial and
tangential VDF's, the velo\-city anisotropy at any given radius can
easily be determined, as we will show later in this section, since it
is just an integral over these distributions, $\sigma^2 = \int v^2
f(v) dv/\int f(v) dv$. The shape of the radial VDF changes as a
function of radius (section 2.1), whereas the shape of the tangential VDF is
virtually constant (section 2.2), and it is therefore natural to expect that the
velocity anisotropy will also change as a function of radius. Since the radial 
VDF generally is more flat-topped than the tangential one (see
fig.~\ref{fig:tangent.log}, top lines), then we will expect $\beta$ to
be positive. Only when the density slope approaches zero (e.g. a
central core) will the radial VDF approach the tangential one, and
hence $\beta \rightarrow 0$ (see fig.~\ref{fig:tangent.log}, bottom
lines).

Let us assume that the radial density profile is given, e.g. by a
truncated NFW profile.  In this case the radial VDF is completely
determined through the Eddington formula. The tangential VDF is given
by eq.~(\ref{eq:tsallis}) and at first sight there are 2 free
parameters, namely the $\sigma$ entering equation~\ref{eq:tsallis},
and then the normalization.  For a given $\sigma$ we can determine the
normalization, since the particle number is conserved when integrating
over the radial or tangential VDF, $\int f_{rad} dv = \rho = \int
f_{tan} dv$. This leaves us only to determine the $\sigma$ entering
eq.~(\ref{eq:tsallis}). One could argue that it most likely is either
$\sigma_r, \sigma_{tan}$ or $\sigma_{tot}$ which should enter here. We
will allow ourselves to be guided by the results of numerical
simulations, and use the average $\sigma_{tot}$, since that gives a
fairly good approximation to all the tangential VDF's from the
simulations discussed above. We will present a more quantitative test
of this in a future paper, but the effect is modest. E.g. for the
truncated NFW profile at the radius where the slope is $-2$, we find
$\beta=0.265$ when using $\sigma = \sigma_{tot}$, and if we instead
use $\sigma = \sigma_r$ ($\sigma_{tan}$) we get $\beta = 0.22 (0.31)$.

It is now straight-forward to calculate the velocity anisotropy,
$\beta$, at any radius for any given density profile with no free
parameters.  Practically we do it iteratively in the following way.
1) First we find the radial VDF, using the Eddington formula. 2) Then
we write the tangential VDF, which is eq.~(\ref{eq:tsallis}) where we
initially use $\sigma = \sigma_r$ (initially assuming $\beta$ to be
very small). 3) Then replace this form at high momenta with the radial
VDF (as described in detail in section 2), and normalized in such a
way that the particle number is conserved (between radial and
tangential). 4) It is now trivial to calculate $\beta$ as an integral
over these distribution functions, and if this is different from the
initial assumption, then re-iterate the entire process with the
calculated $\beta$. This means explicitely that we return to point 2)
with this newly calculated $\beta$, and $\sigma = \sigma_{tot}$.
In practice we repeat until $\beta$ has converged with accuracy $0.01$.


\begin{figure}[thb]
	\centering
	\includegraphics[angle=0,width=0.49\textwidth]{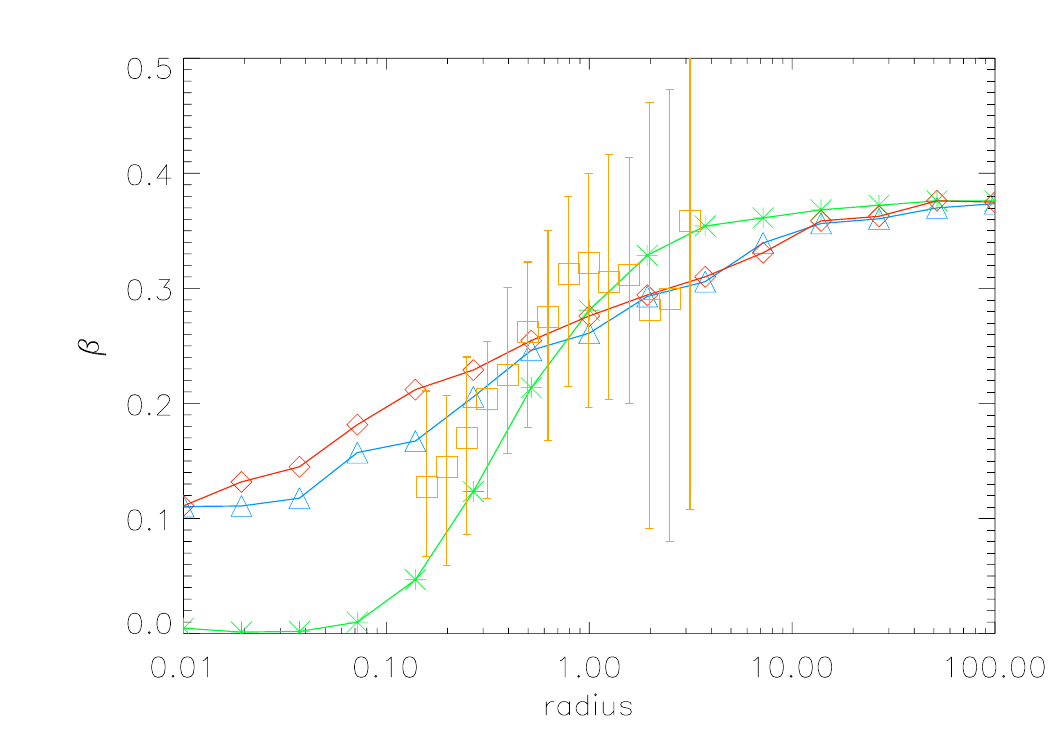}
	\caption{The velocity anisotropy as function of radius.  Blue
triangles are for an NFW density profile, the red diamonds for the density
profile suggested by
\cite{navarro2004}, and the green stars is for $\rho(r) \sim 1/(1+r^2)^2$,
which has a central core (x-axis normalized to the scale radius).  The
squares with error-bars is from the CLEF simulation
\citep{kay,springel}, where the 67 most relaxed galaxy clusters at
$z=0$ have been selected (x-axis normalized to $r_{2500}$). The
error-bars correspond to 1 sigma scatter over the 67 most relaxed
clusters~\citep{host2008}. The $\beta$-profiles from pure dark matter
simulations (e.g. \cite{diemand}) are in good agreement with this
radial behaviour.}
\label{fig:beta.r}
\end{figure}
  

In fig.~\ref{fig:beta.r} we present the radial dependence of $\beta$
for 3 density profiles, namely an NFW profile with a truncation at
large radius, a profile like the one advocated by \cite{navarro2004},
and finally a profile of the form $\rho(r) \sim 1/(1+r^2)^2$, which
has a central core. We see that the anisotropy increases in a way
similar to what is observed in numerical simulations, namely from
something small in the central region, to something of the order 0.4
towards the outer region.  The orange squares are from CLEF numerial
simulation~\citep{kay,springel}, where the error-bars represent the
1$\sigma$ scatter over the 67 most relaxed clusters~\citep{host2008}.
Observations of $\beta (r)$ in galaxy clusters are in excellent
agreement with these numerical predictions~\citep{host2008}.  The
radial scale of the simulated $\beta$ is $r_{2500}$, which does not
have to coincide with the scale radius of the analytical
profiles. This gives a free parameter (of the order unity) in the
normalization of the x-axis, which we just put to 1 for simplicity.
In a similar way the analytical profiles are all normalized to their
respective scale radii, which means that they could also have different
$r_{2500}$.

Since we have suggested the shape of the full VDF's, then we can
naturally also get higher order moments, such as the kurtosis, as
function of radius. We thus also predict that the radial profile of the
higher velocity moments are fully determined by the shape of the
density profile.

\section{Discussion}

One of the consequences of the above considerations is, that the
appearance of the radial variation of $\beta$ is dictated by the
density profile. That is, given any density profile, the velocity
anisotropy is entirely determined, as long as the structure has had
time to equilibrate. We are thus stating explicitly, that $\beta$ is
unrelated to the infall of matter in the outer region, and that 
the only connection $\beta$ has to the formation process is through
the radial structure of the density profile.
This is naturally supported by the very non-cosmological 
simulations (see figs.~\ref{fig:tangent}, \ref{fig:tangent.log})
which also produce a $\beta$-profile in agreement with cosmological
simulations~\citep{hansenzemp}.


\begin{figure}[thb]
	\centering
	\includegraphics[angle=0,width=0.49\textwidth]{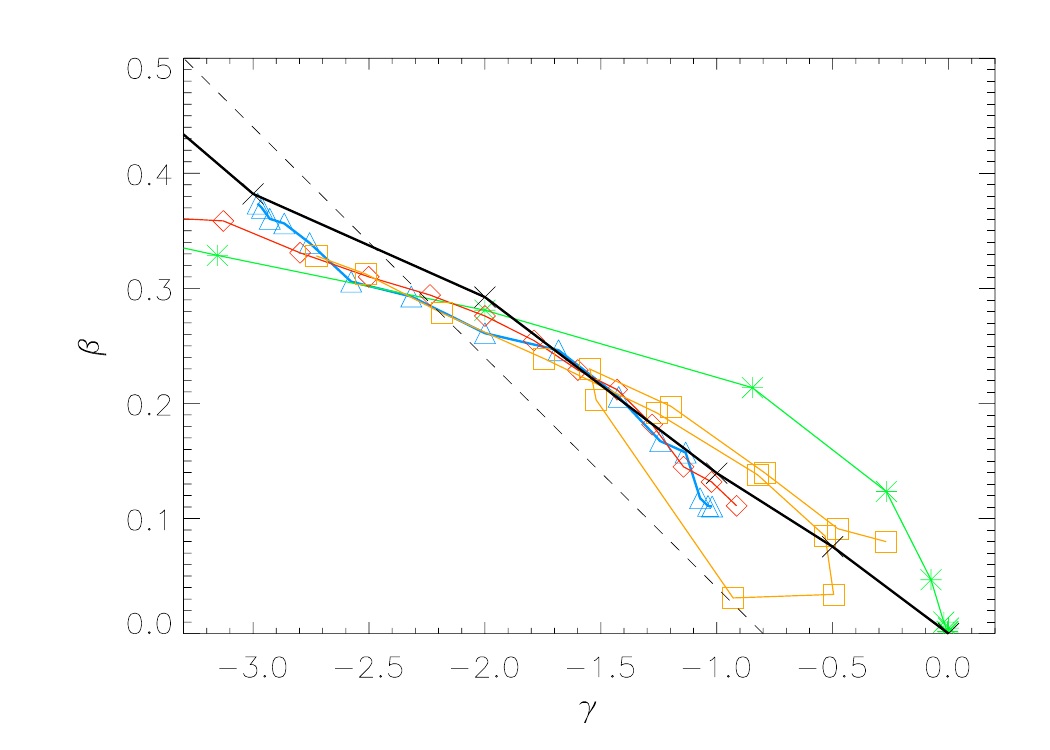}
	\caption{The velocity anisotropy, $\beta$, 
as function of the density slope, $\gamma$. The blue line (triangles)
is for an NFW profile, and the black solid line (crosses) is for a power-law
density profile.  The orange (squares) is for the non-trivial 
double-bump profile, 
the green (stars) is for $\rho = 1/(1+r^2)^2$ profile,
and the red diamonds are for the profile suggested in \cite{navarro2004}.
The dashed straight line is the suggestion from
\cite{hansenstadel}. These results are roughly fit by $\beta = -0.13\gamma$.}
\label{fig:gamma.beta}
\end{figure}


Another consequence is that $\beta$ must always be positive
in equilibrated structures, since the density profile at most
can develop a core (see fig.~\ref{fig:beta.r}).
This is also in good agreement with dark matter simulations
\citep{dehnenmclaughlin,barnes,faltenb,bellovary}.
Virtually no numerical simulations find negative velocity anisotropy, and
when they do, this is usually only in the very inner region where numerical
convergence may be questioned. Few analytical treatments have predicted
a negative $\beta$ in the inner region \citep{zait}, however, this
result may be an artefact of assuming that the pseudo phases-space is a
perfect power-law in radius, which is generally not correct \citep{schmidt09}.
If future high-resolution numerical simulations instead will establish that
the central velocity anisotropy is negative (in agreement with the predictions
of \cite{zait}), then that would be a proof
that the present analysis is flawed somehow.

It has previously been suggested that a connection between the
anisotropy and the slope of the density profile should exist.  This
connection appears to hold even for structures which have profiles
with non-trivial radial variation in $d{\rm log}\rho/d{\rm log}r$
\citep{hansenmoore}.  We can now test this connection. In
figure~\ref{fig:gamma.beta} we show $\beta$ as function of the density
slope for the NFW profile (solid line, blue triangles), the profile
suggested by \cite{navarro2004} (red diamonds), and also for a
``double-bump'' structure (the sum of two spatially separated profiles
of the form $1/(1+r)^3$, orange squares)).  This double-hump profile
has a very non-trivial radial variation of $d{\rm ln}\rho/d{\rm ln}r$,
which cannot be well approximated by any generalized double power-law
profile.  All these structures appear to land near a connection
roughly given by $\beta = -0.13\gamma$. We also show the results for
the $\rho = 1/(1+r^2)^2$ profile (green stars), as well as for single
power-law profiles (fat black line, crosses). The dashes line is
$\beta = -0.2(\gamma-0.8)$ as suggested in \cite{hansenstadel}, based
on a set of cosmological and non-cosmological simulations. 
These two results differ by approximately 0.1 in $\beta$. We
indeed see that all structures land in a relatively narrow band in the
$\gamma$--$\beta$ plane, and hence likely explaining the origin of the
$\gamma$--$\beta$ relations.

The most important practical implication of this suggestion is, that it
will allow us to close the Jeans equation. As is well-known, the Jeans
equation depends on the density, dispersion, anisotropy and the total
mass. Now, having demonstrated (or at least suggested strongly) that
the anisotropy is uniquely determined once the density is known, we
see that it is possible to close the Jeans equation for systems that
are fully relaxed.

We have been making simplifying assumptions above, which all need to
be tested through high resolution simulations.  First, we assume that
the radial VDF is very similar to the one appearing from the
Eddington formula, even in the presence of a non-zero $\beta$.  We
also assume that the $\sigma$ entering eq.~(\ref{eq:tsallis}) is the
total one.  If the correct $\sigma$ to use is instead closer to
$\sigma_{tan}$, then $\beta$ will be slightly larger, but the radial
variation will remain.  From these assumptions we estimate the
accuracy of the present work to be about 0.1 or up to about $30\%$ in
$\beta(r)$.

One could naturally ask why and how the radial and tangential VDF's
get their shapes?  It is slightly disappointing that it is not a deep
physical principle, like a generalized entropy, which is
responsible. Instead it is simply the density profile (either the
radially varying, or the tangentially constant) which through the
Eddington formula demands that the VDF's take on these forms. These
forms will therefore appear when there is sufficient amount of violent
relaxation to allow enough energy exchange between the particles.

\section{Summary}

The velocity of any particle can be decomposed into the radial and
tangential components, and when summing over all particles in a radial
bin, we get the particle velocity distribution function, the VDF. We
suggest that both the radial and tangential VDF's are given through
the Eddington formula. The radial one comes from the radially changing
density profile, and the tangential VDF arises when considering a
structure with constant density and potential.  This is because
the tangential component of the velocity as a first approximation is
moving in constant density and constant potential.  In addition the
tangential VDF is reduced for high-energy particles in accordance with
the radial VDF, to ensure that the particles remain bound to the
structure. These phenomenological predictions are in remarkably good
agreement with the results from numerical simulations of collisionless
particles, both of structures of cosmological origin as well as highly
non-cosmological origin.

Under these suggestions it is straight forward to derive the velocity
anisotropy profile, $\beta (r)$, with no free parameters. This is
shown to increase radially from something small (possibly zero) in the
center, to something large and positive (possibly around 0.4) towards
the outer region.

We have thus demonstrated that the velocity anisotropy is entirely
determined from the density profile.  This allows us to close the
Jeans equation, since $\beta$ is no-longer a free parameter.

\noindent

{\bf Acknowledgements}\\ It is a pleasure to thank 
Jin An and  Ole Host for discussions, and Ben Moore and
Jesper Sommer-Larsen for kindly letting me use their 
simulations.
The Dark Cosmology Centre is funded by the Danish National Research Foundation.




\begin{thebibliography}{99}


\bibitem[Ascasibar \& Gottloeber(2008)]{ascasibar}
Ascasibar, Y., \& Gottloeber, S. 2008, arXiv:0802.4348
  


\bibitem[Austin et al.(2005)]{austin}
Austin, C.~G., Williams, L.~L.~R., Barnes, E.~I., Babul, A., \& 
Dalcanton, J. J. 2006,
ApJ, 634, 756

\bibitem[Barnes et al.(2007)]{barnes}
Barnes, E.~I., Williams, L.~L.~R., Babul, A., \& Dalcanton, J. J.  2007,
Astrophys.\ J.\  {\bf 654} (2007) 814
  
  
  

\bibitem[Bellovary et al.(2008)]{bellovary}
Bellovary, J. M. et al. 2008,
arXiv:0806.3434
  


\bibitem[Binney(1982)]{binney82}
Binney, J. 1982, MNRAS, 200, 951

\bibitem[Binney \& Tremaine(1987)]{binneytremaine}
Binney, J., \& Tremaine, S.\ 1987, Princeton, NJ, 
Princeton University Press, 1987, 747 p.


\bibitem[de Blok et al.(2001)]{blok01}
de Blok, W. J. G., McGaugh, S. S.,  Bosma, A., \& Rubin, V. C. 2001,
ApJ, 552, 23 

\bibitem[de Blok, Bosma \& McGaugh(2003)]{blok02}
de Blok W. J. G., Bosma A., \& McGaugh S. S.
2003, MNRAS, 340, 657


\bibitem[Broadhurst et al.(2005)]{broadhurst}
Broadhurst, T.~J.,
Takada, M., Umetsu, K., Kong, X., Arimoto, N.,
Chiba, M., \& Futamase, T. 2005, Astrophys.\ J.\  619, L143
%
  
  

\bibitem[Buote \& Lewis(2004)]{buote}
Buote, D.~A., \& Lewis, A.~D. 2004, Astrophys.\ J.\   604, 116
  
  


\bibitem[Carlberg et al.(1997)]{carlberg} 
Carlberg, R.~G., et al.\ 1997,
ApJ, 485, L13


\bibitem[Cole \& Lacey(1996)]{colelacey}
Cole, S., \& Lacey, C.\ 1996,
MNRAS, 281, 716





\bibitem[Corbelli(2003)]{corb03} 
Corbelli, E.
2003, MNRAS, 342, 199

\bibitem[Courteau(1997)]{courteau97}
Courteau, S. 1997,
AJ, 114, 2402

\bibitem[Cuddeford(1991)]{cuddeford}
Cuddeford, P.\ 1991, 
\mnras, 253, 414 

\bibitem[Dehnen \& McLaughlin(2005)]{dehnenmclaughlin}
Dehnen, W., \& McLaughlin, D. 2005, MNRAS, 363, 1057



\bibitem[Diemand et al.(2004)]{diemand} 
Diemand, J., Moore, B., \& Stadel, J.\ 2004, MNRAS, 353, 624





\bibitem[Eddington(1916)]{eddington}
Eddington, A. S. 1916, MNRAS, 76, 572

\bibitem[Evans \& An(2006)]{evansan}
Evans, N.~W., \&  An, J.~H. 2006, Phys.\ Rev.\  D., 73, 023524


\bibitem[Fairbairn \& Schwetz(2008)]{fairbairn}
Fairbairn, M., \& Schwetz, T. 2008,
arXiv:0808.0704


\bibitem[Faltenbacher \& Diemand(2006)]{faltenb}
Faltenbacher,A., \& J.~Diemand,
. 2006,
Mon.\ Not.\ Roy.\ Astron.\ Soc.\  369, 1698
  
%
  [arXiv:astro-ph/0602197].




\bibitem[Gilmore et al.(2007)]{gilmore}
Gilmore, G., Wilkinson, M.~I.,  Wyse, R.~F.~G., Kleyna, J.~T., Koch, A., 
Evans, N.~W., \&  Grebel, E.~K. 2007, Astrophys.\ J.\ 663, 948
%
  

\bibitem[Gonz{\'a}lez-Casado et al.(2007)]{gsmh} 
Gonz{\'a}lez-Casado, G., Salvador-Sol{\'e}, E.,
Manrique, A., \& Hansen, S.~H.\ 2007, 
arXiv:astro-ph/0702368

\bibitem[Graham et al.(2006)]{alister} 
Graham, A.~W., Merritt, D., Moore, B., Diemand, J.,
\& Terzi{\'c}, B.\ 2006, \aj, 132, 2701



\bibitem[Hansen et al.(2005)]{zurichstudents}
Hansen, S.~H., Egli, D., Hollenstein, L., Salzmann, C. 2005,
New Astron.\  10, 379
  

\bibitem[Hansen \& Moore(2006)]{hansenmoore}
Hansen S.~H., \& Moore B 2006, New Astron., 11, 333

\bibitem[Hansen et al.(2006)]{hansenzemp} 
Hansen, S.~H., Moore, B., Zemp, M., \& Stadel,
J.\ 2006, Journal of Cosmology and Astro-Particle Physics, 1, 14

\bibitem[Hansen \& Stadel(2006)]{hansenstadel} 
Hansen, S.~H., \& Stadel, J.\ 2006, Journal of
Cosmology and Astro-Particle Physics, 5, 14


\bibitem[Hansen(2004)]{hansenjeans}
Hansen S.~H. 2004 MNRAS, 352, L41


\bibitem[Hansen \& Piffaretti(2007)]{hansenpiff}
Hansen, S.~H., \& Piffaretti, R. 2007,
\aap, 476, L37


\bibitem[Host \& Hansen(2007)]{hosthansen}
Host, O., \& Hansen, S.\ H.\ 2007, JCAP, 0706, 016

\bibitem[Host et al.(2009)]{host2008}
Host, O., Hansen, S. H., Piffaretti, R., Morandi, A., Ettori, S.,
Kay, S. T., \& Valdarnini, R. 2009, ApJ to appear,
arXiv:0808.2049


\bibitem[Henriksen(2007)]{henriksen}
Henriksen, R.~N.\ 2007, arXiv:0709.0434 

\bibitem[Henriksen(2008)]{henriksen2}
Henriksen, R. N.\ 2008, submitted to ApJ

\bibitem[Kay et al.(2007)]{kay}
Kay, S.~T., da Silva, A.~C., Aghanim, N., Blanchard, A., Liddle, A.~R., 
Puget, J.-L., Sadat, R., \& Thomas, P.~A.\ 2007,
\mnras, 377, 317 


\bibitem[Knollmann et al.(2008)]{knollmann}
Knollmann, S. R., Knebe, A. \& Hoffman, Y. 2008,
arXiv:0809.1439


\bibitem[Merritt et al.(2006)]{merritt} 
Merritt, D., Graham, A.~W., Moore, B., Diemand, J.,
\& Terzi{\'c}, B.\ 2006, \aj, 132, 2685

\bibitem[Moore et al.(1998)]{moore}
Moore, B., Governato, F., Quinn, T., Stadel, J. \& Lake G. 1998,
ApJ, 499, 5

 

\bibitem[Moore et al.(2001)]{moore2001} 
Moore, B., Calcaneo-Roldan, C., Stadel, J., Quinn, T.,  Lake, G.,
Ghigna, S., \& Governato, F. 2001,
Phys.\ Rev.\ D., 64, 063508
 
 



\bibitem[Navarro et al.(1996)]{nfw}
Navarro, J. F., Frenk, C. S., \&  White, S. D. M.\
1996, ApJ, 462, 563

\bibitem[Navarro et al.(2004)]{navarro2004}
Navarro, J. et al. 2004,
MNRAS, 349, 1039

\bibitem[Navarro et al.(2008)]{springel08} Navarro, J.~F., et al.\ 
2008, arXiv:0810.1522 

\bibitem[Palunas \& Williams(2000)]{palunas00}
Palunas, P., \& Williams, T. B. 2000,
AJ, 120, 2884

\bibitem[Pointecouteau et al.(2005)]{pointe}
Pointecouteau, E., Arnaud, M., \& Pratt, G.~W. 2005, 
Astron. \& Astrophys, 435, 1
  



\bibitem[Reed et al.(2003)]{reed}
Reed, D. et al. 2003, 
Mon.\ Not.\ Roy.\ Astron.\ Soc.\  357, 82
  

\bibitem[Rubin et al.(1985)]{rubin85}
Rubin, V. C., Burstein, D., Ford, W. K., \& Thonnard, N. 1985,
ApJ, 289, 81





\bibitem[Salvador-Sole et al.(2007)]{smgh07}
Salvador-Sol\'e, E.,
Manrique, A., Gonz\'alez-Casado, G., \& Hansen, S.~H. 2007, 
\apj, 666, 181


\bibitem[Salucci(2001)]{salucci01}
Salucci, P. 2001,
MNRAS, 320L, L1


\bibitem[Salucci et al.(2003)]{salucci}
Salucci, P., Walter, F., \& Borriello, A.\ 2003, \aap, 409, 53 


\bibitem[Salucci et al.(2007)]{salucci2}
Salucci, P., Lapi, A., 
Tonini, C., Gentile, G., Yegorova, I., \& Klein, U.\ 2007, 
\mnras, 378, 41

\bibitem[Sand et al. (2004)]{sand}
Sand, D.~J., Treu, T., Smith, G.~P., \& Ellis, R.~S.
2004, Astrophys.\ J.\  604, 88
  
  



\bibitem[Schmidt et al.(2008)]{schmidt}
Schmidt, K., Hansen, S. H., An, J. H., Williams, L. L. R., \&
Maccio, A. V. 2008, submitted to ApJ

\bibitem[Schmidt et al.(2008)]{schmidt09} Schmidt, K.~B., Hansen, 
S.~H., \& Macci{\`o}, A.~V.\ 2008, \apjl, 689, L33 


\bibitem[Sommer-Larsen (2006)]{sommerlarsen}
Sommer-Larsen, J. 2006, \apj , 644, L1

\bibitem[Sommer-Larsen et al.(2003)]{sommerlarsen2}
Sommer-Larsen, J., G{\"o}tz, M., \& Portinari, L.\ 2003,
\apj, 596, 47 

\bibitem[Springel(2005)]{springel}
Springel, V.\ 2005, \mnras, 364, 1105 

\bibitem[Stadel et al.(2008)]{stadel08} Stadel, J., Potter, D., 
Moore, B., Diemand, J., Madau, P., Zemp, M., Kuhlen, M., 
\& Quilis, V.\ 2008, arXiv:0808.2981 


\bibitem[Stoehr(2004)]{stoehr}
Stoehr, F. 2004, 
Mon.\ Not.\ Roy.\ Astron.\ Soc.\  365, 147

\bibitem[Swaters et al.(2002)]{swaters02}
Swaters, R. A., Madore, B. F., van den Bosch, F. C., \& Balcells, M. 2003,
ApJ, 583, 732



\bibitem[Taylor \& Navarro(2001)]{taylornavarro}
Taylor, J. E., \& Navarro, J. F. 2001,
ApJ, 563, 483

\bibitem[Tsallis(1988)]{tsallis}
Tsallis, C. 1988, J. Stat. Phys., 52, 479


\bibitem[Van Hese et al.(2008)]{vanhese}
Van Hese,  E., Baes, M., \& Dejonghe, H. 2008,
arXiv:0809.0901

\bibitem[Vikhlinin et al.(2006)]{vikhlinin}
Vikhlinin, A., 
Kravtsov, A., Forman, W., Jones, C., Markevitch, M.,
Murray, S.~S., \& Van Speybroeck, L.
2006, Astrophys.\ J.\  640, 691
  
  

\bibitem[Wilkinson et al.(2004)]{wilkinson}
Wilkinson, M.~I. et al. 2004, Astrophys.\ J.\  611, L21
  
  

\bibitem[Wojtak et al.(2005)]{wojtak2005}
Wojtak, R. et al. 
Lokas, E.~L., Gottloeber,  S., \& Mamon, G.~A.
2005, Mon.\ Not.\ Roy.\ Astron.\ Soc.\  361, L1

\bibitem[Wojtak et al.(2008)]{wojtak2008}
Wojtak, R., Lokas, E.~L., Mamon, G.~A.,
Gottloeber, S., Klypin, A., \&  Hoffman, Y.
2008, arXiv:0802.0429

\bibitem[Zait et al.(2008)]{zait}
Zait, A., Hoffman, Y., \&  Shlosman, I. 2008,
arXiv:0711.3791




\end{thebibliography}
\end{document}